\documentclass[%
 reprint,
superscriptaddress,
 amsmath,amssymb,
 aps,
prd,
longbibliography
]{revtex4-1}

\usepackage{aas_macros}
\usepackage{graphicx}
\usepackage{dcolumn}
\usepackage{bm}
\usepackage{hyperref}

\usepackage{color}
\usepackage[mathlines]{lineno}
\usepackage{tabularx}
\usepackage{subfigure}
\usepackage[normalem]{ulem}
\usepackage[export]{adjustbox}
\usepackage{url}
\usepackage{academicons}
\usepackage{xcolor}
\usepackage{orcidlink}

\hypersetup{
    pdfnewwindow=true,    
    colorlinks=true,      
    linkcolor=blue,       
    citecolor=blue,       
    filecolor=blue,       
    urlcolor=blue         
}

\usepackage{natbib}

\begin{document}
\title{Sensitivity floor for 
primordial black holes with neutrino searches}

\author{Qishan Liu \orcidlink{0000-0003-1437-6829}}
\email{qsliu@phy.cuhk.edu.hk}
\thanks{\scriptsize \!\! \href{https://orcid.org/0000-0003-1437-6829}{orcid.org/0000-0003-1437-6829}}
\affiliation{Department of Physics, The Chinese University of Hong Kong, Shatin, New Territories, Hong Kong}

\author{Kenny C. Y. Ng \orcidlink{0000-0001-8016-2170}}
\email{kcyng@cuhk.edu.hk}
\thanks{\scriptsize \!\! \href{http://orcid.org/0000-0001-8016-2170}{orcid.org/0000-0001-8016-2170}}
\affiliation{Department of Physics, The Chinese University of Hong Kong, Shatin, New Territories, Hong Kong}

\date{September 24, 2024}

\begin{abstract}
Primordial black holes~(PBHs) formed in the early Universe are well-motivated dark matter~(DM) candidates over a wide range of masses.  These PBHs could emit detectable signals in the form of photons, electrons, and neutrinos through Hawking radiation. 
We consider the null observations of astrophysical $\bar{\nu}_{e}$ flux from several neutrino detectors and set new constraints on the PBHs as the dominant DM component to be above $6.4\times10^{15}\,{\rm g}$. 
We also estimate the expected constraints with JUNO for the prospects in the near future.  Lastly, we note that the diffuse supernova neutrino background~(DSNB) is an unavoidable isotropic background. We thus estimate the sensitivity floor for PBH parameter space due to the DSNB and show that it is challenging for neutrino detectors to identify PBHs as they constitute 100\% of the DM above a mass of 9$\times10^{15}$g. 
\end{abstract}
\maketitle

\section{Introduction} \label{sec:introduction}

Dark matter~(DM) is a major component of the Universe, making up about $26\%$ of the total mass-energy content according to the standard Lambda-CDM model~\cite{ParticleDataGroup:2022pth}.  One of the DM potential candidates is primordial black holes~(PBHs).
During the early stages of the Universe, a range of mechanisms could  lead to the formation of PBHs. These mechanisms include the gravitational collapse of overdense regions arising from primordial inhomogeneities, collisions of cosmic bubbles, the collapse of cosmic loops, the collapse of domain walls, etc~\cite{Carr:1974nx,Hawking:1971ei,Belotsky:2014kca,Khlopov:2008qy,Carr:2009jm,Carr:2020gox}.

If PBHs were formed and survived until the present time, they could represent all or part of the DM population.
While no detection of PBHs has been made thus far, various considerations based on PBHs have been considered to place constraints on the physics governing the early Universe~\cite{Page:1976wx,Josan:2009qn,Kalaja:2019uju, Gow:2020bzo}.

Recently, there has been a renewed interest in PBHs as a DM candidate following the detection of gravitational wave events by LIGO~\cite{LIGOScientific:2016aoc}. 
These events could potentially be attributed to PBH mergers~\cite{Sasaki:2016jop}. PBHs exhibit a broad range of masses, extending from the Planck mass to several hundred times the mass of the Sun, and their specific mass depends on the time of their formation~\cite{Carr:2009jm}.

Numerous studies have explored constraints on the abundance of PBHs across a wide range of masses~\cite{Green:2020jor}. For larger PBHs masses, various gravitational observations can be utilized to impose constraints, including microlensing, accretion, discreteness effects, and gravitational wave observations. For a comprehensive overview of these constraints, see e.g., Refs~\cite{Green:2020jor,Carr:2020gox}.

Light PBHs with masses above $10^{14}\,{\rm g}$ can potentially be detected indirectly as they emit detectable particles and evaporate through Hawking radiation~\cite{Hawking:1974rv,Hawking:1975vcx,Page:1976df}.  
For PBHs with mass smaller than that, they cannot be a viable DM candidate as they would have evaporated by now. 
Only nonrotating (rotating) PBHs with masses higher than $5 \times 10^{14}\,{\rm g}$ ($7 \times 10^{14}\,{\rm g}$) could survive until today~\cite{Page:1976ki,Page:1977um}.

Constraints on the abundance of PBHs with masses ranging from $10^{14}$ g to $10^{17}$\,g have been placed by considering Hawking radiation~\cite{Auffinger:2022khh} in various astrophysical observations. These include galactic gamma rays with INTEGRAL~\cite{Laha:2020ivk}, extra-galactic X-ray and gamma-ray background~\cite{Ballesteros:2019exr,Arbey:2019vqx,Coogan:2020tuf,Carr:2009jm,Ray:2021mxu}, reionization of the early Universe in global 21-cm measurements~\cite{Mittal:2021egv,Saha:2021pqf}, imprints on the cosmic microwave background~\cite{Clark:2016nst,Acharya:2020jbv}, and electrons/positrons emission by considering the Galactic Center 511 keV gamma-ray line~\cite{Dasgupta:2019cae,Laha:2019ssq}.

The potential to search for PBH neutrinos has been considered for a long time~\cite{Halzen:1995hu,Bugaev:2000bz}. More recently, upper limits on the $\bar{\nu}_{e}$ from neutrino experiments can also be used to place constraints on the fraction of PBHs in DM searches~\cite{Wang:2020uvi,Dasgupta:2019cae,Bernal:2022swt, DeRomeri:2021xgy}. Additionally, prospective bounds from the LZ/XENONnT and DARWIN experiment via coherent elastic neutrino-nucleus scattering~\cite{Calabrese:2021zfq} and detection prospects at IceCube~\cite{Capanema:2021hnm} have been explored.
Although the limits are generally weaker, neutrino searches are more robust with respect to DM density profiles due to the all-sky nature of the search, as well as completely different systematics associated with propagation of electromagnetic messengers~\cite{Capanema:2021hnm}. 

There are current upper limits on the $\bar{\nu}_{e}$ from various experiments, such as SNO~\cite{SNO:2004eru}, Borexino~\cite{Borexino:2019wln}, Super-Kamiokande~(SK)~\cite{Super-Kamiokande:2011lwo,Super-Kamiokande:2013ufi,Super-Kamiokande:2021jaq}, and KamLAND~\cite{KamLAND:2021gvi}. 
Future detectors are expected to further improve the sensitivities, such as DUNE~\cite{Moller:2018kpn,DeRomeri:2021xgy}, JUNO~\cite{JUNO:2015zny}, THEIA~\cite{DeRomeri:2021xgy}, the gadolinium phase of Super-Kamiokande (SK-Gd)~\cite{Super-Kamiokande:2023xup,FernandezMenendez:2017ccn}, and Hyper-Kamiokande (HK)~\cite{Moller:2018kpn}. 
While these detectors will improve the PBH sensitivities (or any new sources of MeV neutrinos), they are likely also going to detect the diffuse supernova neutrino background~(DSNB)~\cite{Beacom:2010kk,Lunardini:2010ab, Ando:2023fcc} soon. 
The DSNB will then become a problematic background for PBH $\bar{\nu}_{e}$ searches. 

In this work, we set new constraints on PBHs by considering an array of existing upper limits on $\bar{\nu}_{e}$ flux, we then consider the prospects of future detectors for probing PBHs with neutrinos, such as JUNO~\cite{JUNO:2022lpc}. Finally, we also estimate the sensitivity ``floor'' caused by the DSNB.

\section{neutrino flux from pbhs}

PBHs could have formed with a broad range of masses as a result of the gravitational collapse of overdensities in the early Universe. 
If PBHs are present in the current Universe, they would contribute to the overall DM content. We denote the abundance of PBHs as a fraction of the DM density, which can be represented by,
\begin{equation}
f_{\mathrm{PBH}} \equiv \frac{\Omega_{\mathrm{PBH}}}{\Omega_{\mathrm{DM}}},
\end{equation}
where $\Omega_{\mathrm{PBH}}$ and $\Omega_{\mathrm{DM}}$ are PBHs and DM density fractions, respectively.

At the event horizon of a black hole, quantum fluctuations can create particle-antiparticle pairs, which is known as Hawking radiation~\cite{Hawking:1974rv,Hawking:1975vcx,Page:1976df}. As a result, PBHs can evaporate as the created particles/antiparticles escape, causing a net flux of radiation and a loss of mass energy for the black hole. 
The lifetime of a black hole is approximately given by~\cite{MacGibbon:1991tj,MacGibbon:2007yq},
\begin{equation}\label{eq:time}
t_{\text {evap }}= t_{\text {Universe }}\left(\frac{M_{\mathrm{PBH}}}{5\times 10^{14}g} \right)^{3} ,
\end{equation}
where $M_{\mathrm{PBH}}$ is the PBH initial mass, and $t_{\text {Universe}} \approx 13.8$ Gyr is the age of the Universe. We can infer from Eq.~(\ref{eq:time}) that PBHs weighing less than $5 \times 10^{14}\,{\rm g}$ would have evaporated by now. Thus, our analysis is focused on PBHs with initial masses greater than this threshold. This ensures that the PBHs we consider could still exist in the present day and have the potential to contribute to DM.

The Hawking radiation emits particle species $i$ that exhibits a nearly black-body spectrum, especially in the high-energy limit. 
At lower energy,  when the wavelengths of the particles are comparable to the size of black hole, the emitted particles can be absorbed or scattered by the black hole itself, leading to a suppression of the emission rate~\cite{Page:1976df,Parker:1975jm}. 
This causes the spectrum to deviate from a blackbody, and is encapsulated in the graybody factor $\Gamma_i$.
Thus, the graybody factor quantifies the probability of these emitted particles escaping from the black hole event horizon~\cite{Page:1976df,MacGibbon:1991tj,MacGibbon:1990zk}.
The emitted particles spectrum rate is then
\begin{equation}
\left.\frac{d^2 N_i(E, t)}{ d E dt}\right|_{\text {prim }}=\frac{g_i}{2 \pi} \frac{\Gamma_i\left(E, M_{\mathrm{PBH}}\right)}{e^{E / T_{\mathrm{BH}}} \pm 1},
\end{equation}
where $g_i$ is a particle degrees of freedom, and $-1$ and $+1$ in the denominator refer to bosons or fermions for the final states, respectively. $T_{\mathrm{BH}}$ is the temperature of a PBH given by~\cite{Hawking:1975vcx,MacGibbon:1990zk}
\begin{equation}
T_{\mathrm{BH}}=\frac{1}{4 \pi G M_{\mathrm{PBH}}} \frac{\sqrt{1-a_*^2}}{1+\sqrt{1-a_*^2}},
\end{equation}
where $G$ is the gravitational constant. The reduced spin parameter of a PBH is given by $a^* = J / M_{\mathrm{PBH}}^2$, where $J$ is the angular momentum of the PBH. For nonrotating PBHs, we take $J = 0$, resulting in a black hole temperature of $ T_{\mathrm{BH}} = 1.06 \times \left(\frac{10^{13} \mathrm{~g}}{M_{\mathrm{PBH}}}\right)\mathrm{GeV}$. The peak energy of the neutrino flux is approximately given by $ {E_\nu}\simeq 4.02 \times T_{\mathrm{BH}} $~\cite{MacGibbon:2007yq}. As PBHs evaporate, their mass decreases, leading to an increase in temperature over a time scale roughly the PBH evaporation time scale $t_{\text {evap }}$. As we are mostly interested in cosmological stable PBH DM with a relative large mass~($10^{15}-10^{16}$\,g), this effect can be safely ignored~\cite{Tabasi:2021cxo,Arbey:2021mbl}.

Particles emitted by Hawking radiation could also undergo secondary processes, such as hadronization and decays. These processes thus also contribute to the production of stable messenger, such as photons and neutrinos. 
To calculate both the primary and secondary neutrino fluxes for each PBH mass, we use the BlackHawk v1.2 code.  This version is used instead of the latest v2.1 because the Python package Hazma~\cite{Coogan:2019qpu} used in the v2.1 does not provide neutrino spectra, yet~\cite{Arbey:2021mbl,Arbey:2019mbc}. 
The total neutrino spectrum thus include both the primary and secondary components,
\begin{equation}
\frac{d^{2} N_{\nu}(E_{\nu},t)}{d E_{\nu} d t} = \left.\frac{d^{2} N_{\nu}(E_{\nu},t)}{d E_{\nu} d t}\right|_{\text {prim }} + \left.\frac{d^{2} N_{\nu}(E_{\nu},t)}{d E_{\nu} d t}\right|_{\text {sec }}.
\end{equation}
The spectrum of $\bar{\nu}_{e}$ is obtained by dividing the total spectrum of electron neutrinos by 2~\cite{Lunardini:2019zob}. We do not take into account neutrino oscillations, as they would only have a minor impact on the low-energy neutrino fluxes, resulting in a deviation of less than 2.5\%~\cite{Wang:2020uvi}.

In this work, we consider PBHs with monochromatic mass distributions to calculate the Hawking radiation. For PBHs with extended mass distributions, it generally would lead to a more stringent constraint as constraints from lighter PBHs are typically much stronger~\cite{Dasgupta:2019cae}.

We consider contributions from both the galactic and extra-galactic PBH DM evaporation.  The all-sky extra-galactic flux is
\begin{equation}
\begin{aligned}
\frac{d \Phi_{\mathrm{EG}}}{d E_{\nu}} = 
\frac{f_{\mathrm{PBH}}\rho_{\mathrm{DM}}}{M_{\mathrm{PBH}}} \int_{0}^{z_{\max }} \frac{cd z}{H(z)} \frac{d^{2} N_{\nu}((1+z)E_{\nu},t)}{d E_{\nu} d t},
\end{aligned}\label{eq:EG}
\end{equation}
where $z$ is the redshift, and $c$ is the speed of light. $\rho_{\mathrm{DM}} =  \rho_c \Omega_{\mathrm{DM}}= 1.3 \times 10^{-6} \mathrm{~GeV} \mathrm{~cm}^{-3}$ is the average DM density of the Universe, where the critical density is $\rho_c = \frac{3H^2_0}{8\pi G}$ and $ \Omega_{\mathrm{DM}} = 0.27$. 
The Hubble expansion rate is $H(z) = H_{0} \sqrt{\Omega_{r}(1+z)^4+\Omega_{M}(1+z)^3+\Omega_{\Lambda}}$, where $H_{0}=67.4 \mathrm{~km/s/Mpc}$
is the present Hubble expansion rate, $\Omega_{r}$ is the radiation energy density, which is so small that we neglect it here, $\Omega_{M}=0.315$ is the matter energy density, and $\Omega_{\Lambda}=0.685$ is the dark energy density~\cite{Planck:2018vyg}. 
The flux is obtained by integrating from $z=0$ up to when the PBHs were formed~\cite{Ray:2021mxu}. In practice, it is sufficient to integrate up to redshifts of a few.

The all-sky galactic flux from PBHs in the Milky Way is
\begin{equation}
\begin{aligned}
\frac{d \Phi_{\mathrm{MW}}}{d E_{\nu}}=\frac{d^{2} N_{\nu}}{d E_{\nu} d t} \frac{f_{\mathrm{PBH}}}{{M_{\mathrm{PBH}}}} \int \frac{d \Omega}{4 \pi} \int_{0}^{\ell_{\max }} d \ell \rho_{\mathrm{NFW}}(r(\ell, \phi)),
\end{aligned}\label{eq:MW}
\end{equation}
where $\Omega$ is the solid angle, $r(\ell, \phi)=\sqrt{R_{\odot}^{2}+\ell^{2}-2 R_{\odot} \ell \cos \phi}$ is the radial coordinate, $R_{\odot} \simeq 8.5  \,{\rm kpc}$ is the distance from the Sun to the center of the Milky Way, and $\ell$ is the line-of-sight distance. The line-of-sight integral extends up to $\ell_{\max }=R_{\odot} \cos \phi+\sqrt{R_{h}^{2}-R_{\odot}^{2} \sin ^{2} \phi}$, with $R_{h} \simeq 200 \,{\rm kpc}$ being the DM halo virial radius. To approximate the DM density profile of the Milky Way halo, we employ the Navarro-Frenk-White (NFW) profile~\cite{Navarro:1995iw,Navarro:1996gj}, 
\begin{equation}
\rho_{\mathrm{NFW}}(r)=\rho_{\odot}\left(\frac{r}{R_{\odot}}\right)^{-1}\left(\frac{1+R_{\odot} / R_{s}}{1+r / R_{s}}\right)^{2} ,
\end{equation}
where we use $\rho_{\odot}=0.4~ \mathrm{GeV} \mathrm{cm}^{-3}$ for the local DM mass density\cite{Benito:2019ngh,deSalas:2020hbh}, and $R_{s}=20\,{\rm kpc}$ for the scale radius of the Milky Way. The integral of the DM density profile of the whole Milky Way halo is about
\begin{equation}
\int \frac{d \Omega}{4 \pi} \int_{0}^{\ell_{\max }} d \ell \rho_{\mathrm{NFW}}(r(\ell, \phi)) = 2.2 \times 10^{22} \,{\rm GeV cm^{-2}}. 
\end{equation}

\begin{figure}[t!]
    \centering
    \includegraphics[width=3.5in,height=3.0in]{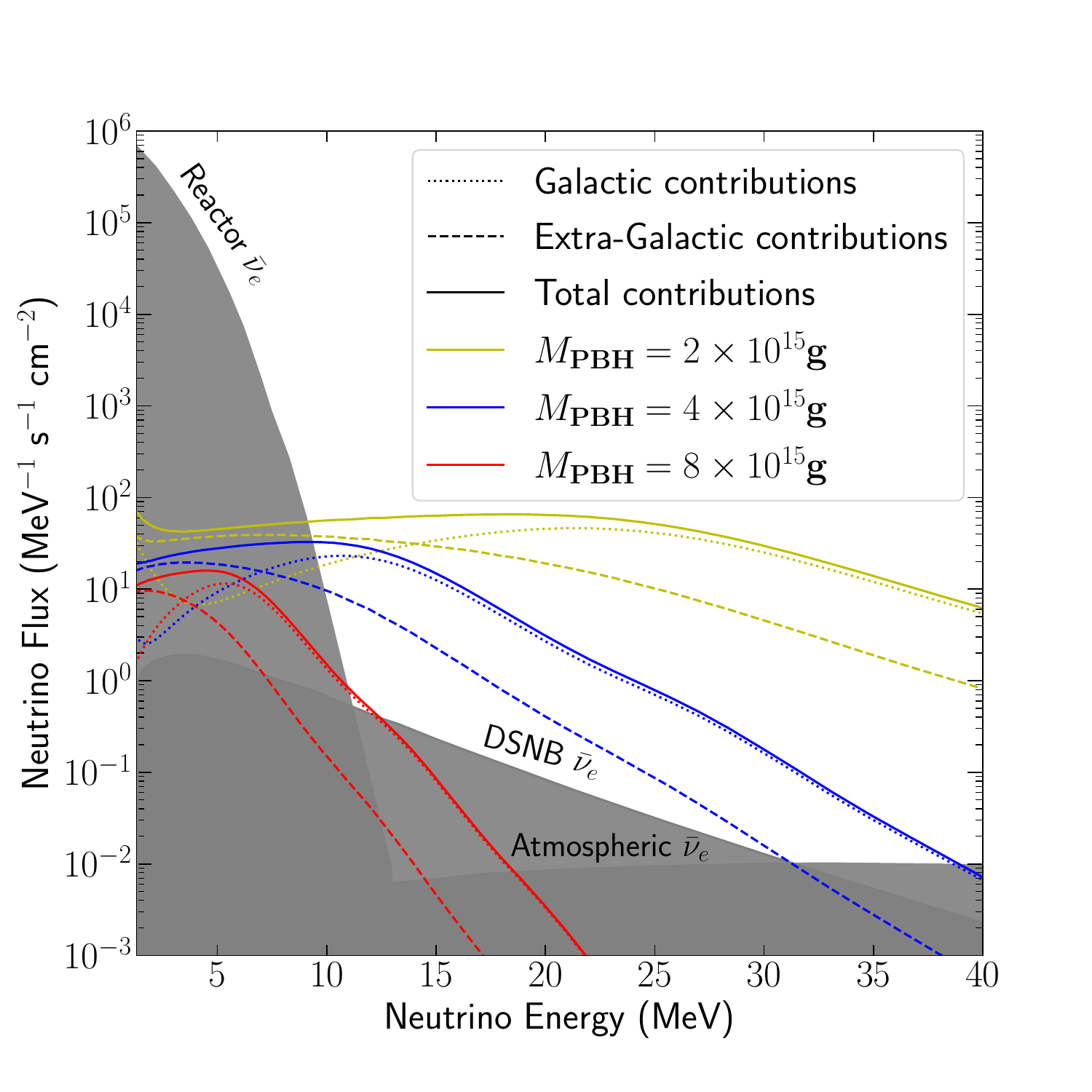}
    \caption{The $\bar{\nu}_{e}$  fluxes  from  PBHs  evaporation as  a  function  of  energy, assuming the PBH fraction $f_{\mathrm{PBH}}= 1$ and a monochromatic mass distribution with masses of $2 \times 10^{15}\,{\rm g}$, $4 \times 10^{15}\,{\rm g}$, and $8 \times 10^{15}\,{\rm g}$. The galactic  (dotted  lines),  extra-galactic  (dashed  lines),  and  total  (solid  lines)  contributions are shown here. The gray shaded regions indicate the most relevant $\bar{\nu}_{e}$ backgrounds at this energy range, which are the reactor~\cite{Wurm:2007cy}, DSNB~\cite{Moller:2018kpn}, and atmospheric $\bar{\nu}_{e}$ fluxes~\cite{Battistoni:2005yu} as labeled. It should be noted that, we show the background at the SK site, but the  reactor $\bar{\nu}_{e}$ backgrounds could be different at another site.}
    \label{fig:bg}
\end{figure}

Fig.~\ref{fig:bg} shows the fluxes of $\bar{\nu}_{e}$ originating from the evaporation of PBHs, as a function of energy. We show the cases of PBHs with masses of $2 \times 10^{15}\,{\rm g}$, $4 \times 10^{15}\,{\rm g}$, and $8 \times 10^{15}\,{\rm g}$, assuming a PBH abundance of $f_{\mathrm{PBH}}= 1$. The relevant backgrounds for $\bar{\nu}_{e}$ in this context include the reactor flux~\cite{Hyper-Kamiokande:2018ofw} (at the SK site), the DSNB~\cite{Moller:2018kpn}, and the atmospheric $\bar{\nu}_{e}$ flux~\cite{Battistoni:2005yu}. 
At low energies, the reactor $\bar{\nu}_{e}$ flux dominates the background, while the DSNB becomes significant in the energy range of $E_{\nu} = 10\,{\rm MeV}$ to $30\,{\rm MeV}$, making it crucial for the search of relatively higher mass PBHs. The atmospheric $\bar{\nu}_{e}$ flux covers the entire energy range but remains subdominant below $30\,{\rm MeV}$.

\section{Current Experimental Upper limits for \texorpdfstring{$\bar{\nu}_{e}$}~ Fluxes }
\subsection{ \texorpdfstring{$\bar{\nu}_{e}$}~ as PBHs probe}

The upper limits on model-independent $\bar{\nu}_{e}$ fluxes can be utilized to constraint on DM fraction of PBHs. There is a low background window for neutrino energy $E_{\nu}$ between roughly $10$ to $30\,{\rm MeV}$ for detecting astrophysical sources of $\bar{\nu}_{e}$, such as PBHs and the DSNB, as shown in Fig.~\ref{fig:bg}. 
 This window is optimal because neutrinos with energies below $10\,{\rm MeV}$ are predominantly affected by reactor neutrino backgrounds~\cite{Wurm:2007cy}, while at higher energies, the background from atmospheric neutrinos~\cite{Battistoni:2005yu} rises rapidly.

The most promising reaction for $\bar{\nu}_{e}$ detection is the inverse beta decay (IBD), $\bar{\nu}_{e}+p \rightarrow n+e^{+}$, which has a large cross section and produces an easily identifiable final state positron. 
In IBD, the energy of the neutrino can be determined by $E_\nu \simeq E_e + 1.8 \,{\rm MeV}$, where the $1.8 \,{\rm MeV}$ accounts for the electron mass and the mass difference between a proton and a neutron. The positron can be detected via Cherenkov or scintillation signals.

Fig.~\ref{fig:data} shows the upper limits on the $\bar{\nu}_{e}$ fluxes obtained by various experiments, including SNO~\cite{SNO:2004eru}, Borexino~\cite{Borexino:2019wln}, SK~\cite{Super-Kamiokande:2011lwo,Super-Kamiokande:2013ufi,Super-Kamiokande:2021jaq}, and KamLAND~\cite{KamLAND:2021gvi}. 
In addition to the experimental upper limits, we also display the $\bar{\nu}_{e}$ fluxes resulting from the evaporation of PBHs with masses of $2 \times 10^{15}\,{\rm g}$, $4 \times 10^{15}\,{\rm g}$, and $8 \times 10^{15}\,{\rm g}$, same as those in Fig.~\ref{fig:bg}.
For comparison, we  include a theoretical prediction of the DSNB flux~\cite{Moller:2018kpn}, which could be new problematic background to the PBH search in the future~(see Sec.~\ref{sec:floors} for details).

JUNO will be the largest liquid scintillator detector for neutrino physics~\cite{JUNO:2015zny} in the near future. It has excellent capabilities for $\bar{\nu}_{e}$ tagging and background rejection, which could strongly improve the $\bar{\nu}_{e}$ flux upper limits. 
We also show the theoretically expected $\bar{\nu}_{e}$ upper limits by JUNO in Fig.~\ref{fig:data}~\cite{JUNO:2022lpc}.

\subsection{ Statistical formalism for \texorpdfstring{$\bar{\nu}_{e}$}~ flux upper limits}\label{sec:chisq}
\begin{figure}[t!]
    \centering
    \includegraphics[width=3.5in,height=3.0in]{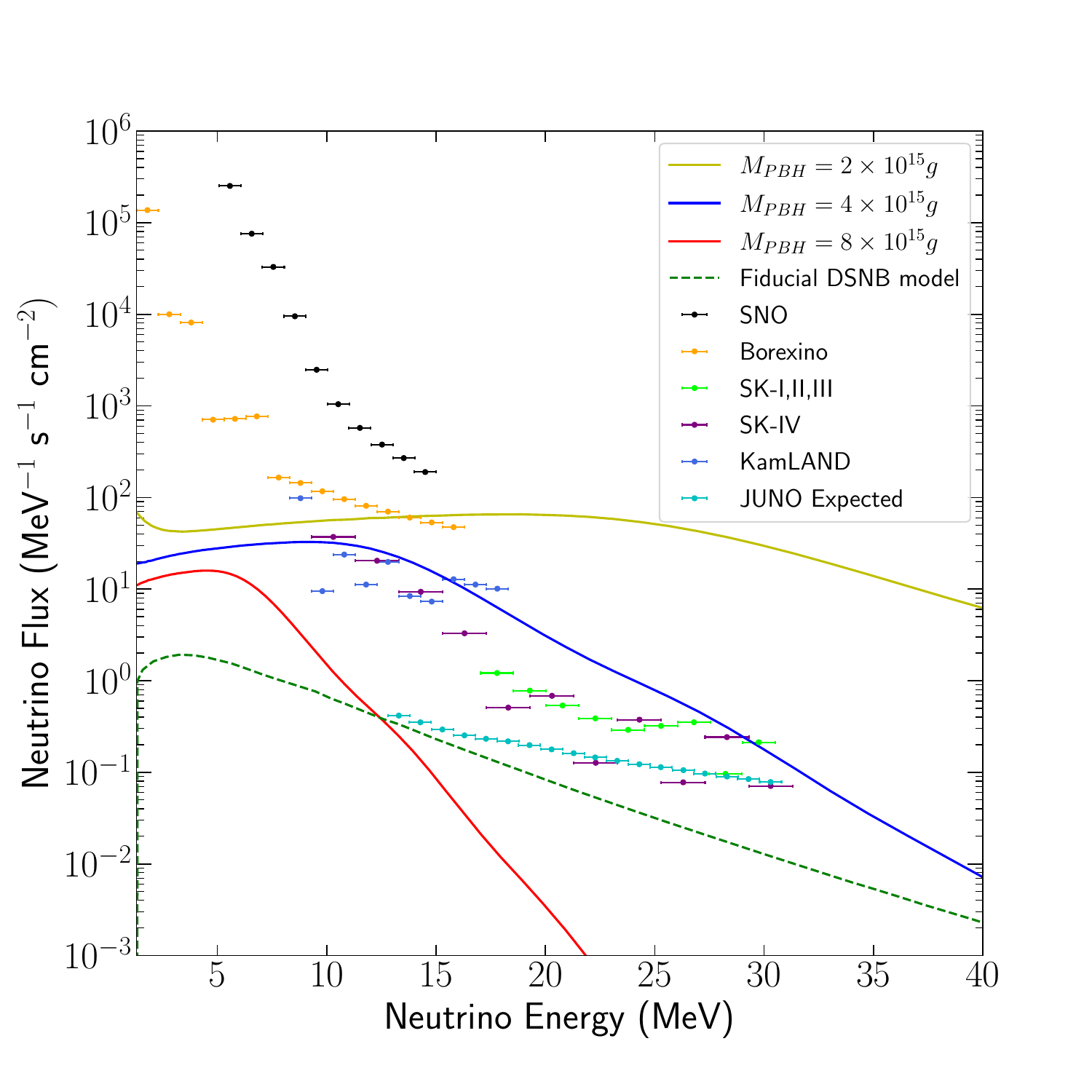}
    \caption{The 90\% confidence level observed upper limits on the $\bar{\nu}_{e}$ flux from different experiments, SNO~\cite{SNO:2004eru}, Borexino~\cite{Borexino:2019wln}, SK~\cite{Super-Kamiokande:2011lwo,Super-Kamiokande:2013ufi,Super-Kamiokande:2021jaq}, and KamLAND~\cite{KamLAND:2021gvi}, as well as the expected upper limits from the upcoming JUNO experiment~\cite{JUNO:2022lpc}. The DSNB theoretical predictions~\cite{Moller:2018kpn}~(dashed lines) and  total  contributions from different PBHs mass~(solid  lines) are shown here, assuming $f_{\mathrm{PBH}}= 1$, and  monochromatic mass distribution with $2 \times 10^{15}\,{\rm g}$, $4 \times 10^{15}\,{\rm g}$ and $8 \times 10^{15}\,{\rm g}$.  }
    \label{fig:data}
\end{figure}

We consider the $\bar{\nu}_{e}$ flux upper limits from various detectors as shown in Fig.~\ref{fig:data}, and perform a simple chi-square analysis to find the corresponding upper limits for PBHs. The chi-square statistic $\chi^{2}$ for each energy bin $i$ is
\begin{equation}\label{eq:chisq}
\begin{aligned}
\chi^{2}_{i}&=\frac{(F_{i} + F^{B}_{i} - F^{obs}_{i})^{2}}{\left(\sigma_{i}\right)^{2}} \simeq \frac{(F_{i})^{2}}{\left(\sigma_{i}\right)^{2}}\, ,
\end{aligned}
\end{equation}
where $F_{i}$ is the PBH model flux, $F^{B}_{i}$ is the background flux of the experiment,  $F_{i}^{obs}$ is the observed flux, and $\sigma_{i}$ is the uncertainty. 
Here we assume that the reported experimental upper limits are obtained when the $F^{B}_{i} \simeq F_{i}^{obs}$.  In other words, there are no significant deviations of the observed data from the expected background. 
In this case, the reported upper limits and the uncertainties are simply related by $\sigma_{i} = F^{up}_{i}/\sqrt{2.71}$, corresponding to the case of $\chi^{2}_{i} = 2.71$ for each bin. 
By substituting this relation into Eq.~(\ref{eq:chisq}), we can construct a general chi-square statistic to test arbitrary models, as prescribed in Ref.~\cite{KamLAND:2011bnd} for KamLAND data.

At each PBH mass, the only parameter of interest is the PBH fraction $f_{\mathrm{PBH}}$. 
The differential $\bar{\nu}_{e}$ fluxes from PBHs are the sum of the galactic and extra-galactic components from Eqs.~(\ref{eq:EG}) and (\ref{eq:MW}),
\begin{equation}
    \frac{d \Phi\left(E_{\nu}\right)}{d E_{\nu}}= \left[ \frac{d \Phi_{\mathrm{EG}}\left(E_{\nu}\right)}{d E_{\nu}} + \frac{d \Phi_{\mathrm{MW}}\left(E_{\nu}\right)}{d E_{\nu}}\right].
\end{equation}

To compare with experimental data, the neutrino flux is binned according to that of data, 
\begin{eqnarray}\label{eq:flux1}
\nonumber F_{i}^{\mathrm{PBH}}\left(f_{\mathrm{PBH}}\right) &=& \frac{\int_{E_{ i}^{\min }}^{E_{ i}^{\max }} \frac{d \Phi\left(E_{\nu}\right)}{d E_{\nu}} \sigma_i(E_{\nu}) d E_{\nu} }{\int_{E_{ i}^{\min }}^{E_{ i}^{\max }}\sigma_i(E_{\nu})d E_{\nu}} \\
&=& f_{\mathrm{PBH}} F_{i}^{\mathrm{PBH}}\left(1\right).
\end{eqnarray}
where $E_{i}^{\min }$ and $E_{i}^{\max }$ are the edge values of an energy bin, $ \sigma_i(E_{\nu})$ corresponds to the IBD cross section~\cite{Strumia:2003zx,Ricciardi:2022pru,Ricciardi:2023tvg} and the abundance factor $f_{\mathrm{PBH}}$ can be factored out. 

Combining Eqs.~(\ref{eq:chisq}) and (\ref{eq:flux1}), the summed chi-square statistics of a set of data from a detector is then 
\begin{equation}\label{eq:flux}
\begin{aligned}
\chi^{2}&=
\sum_{i} \frac{(f_{\mathrm{PBH}} F_{i}^{\mathrm{PBH}}\left(1\right))^{2}}{\left(F^{up}_{i} / \sqrt{2.71}\right)^{2}}.
\end{aligned}
\end{equation}
The one-sided 95\% upper limit on the PBH fraction, $f_{\mathrm{PBH}}^{95}$, can then be obtained by $\chi^{2} =2.71$, or 
\begin{equation}\label{eq:bound}
f_{\mathrm{PBH}}^{95} = \sqrt{\frac{1} {\sum\limits_{i}
\frac{
( F_{i}^{\mathrm{PBH}}\left(1\right))^{2}}{
(F_{i}^{up})^{2}
} }}.
\end{equation}

\subsection{Constraints on PBHs by \texorpdfstring{$\bar{\nu}_{e}$}~ flux upper limits} 

\begin{figure}[t!]
    \centering
    \includegraphics[width=3.5in,height=3.0in]{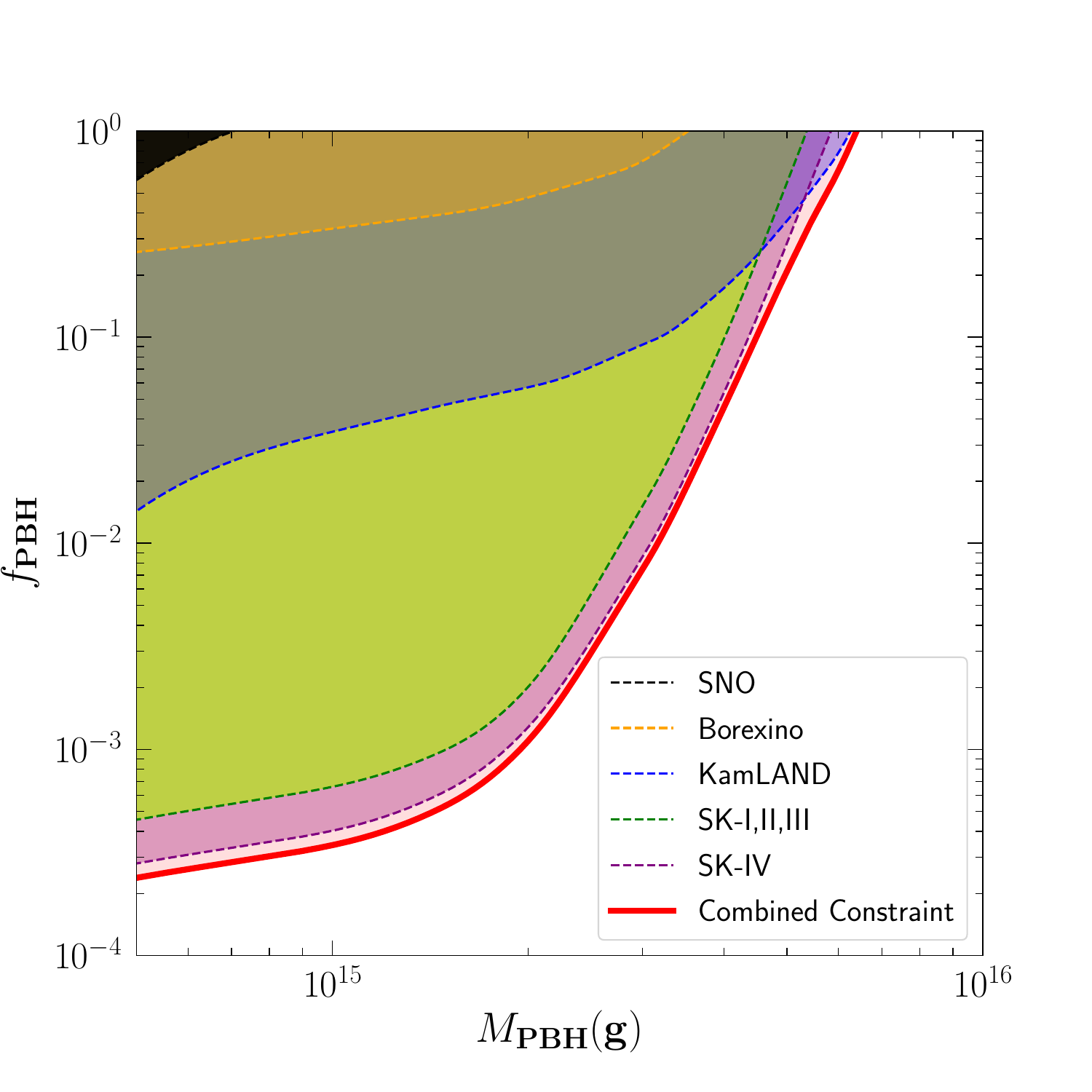}
    \caption{Upper limits on DM fraction of PBHs $f_{\mathrm{PBH}}$, from $\bar{\nu}_{e}$ fluxes at different experiments, SNO~\cite{SNO:2004eru}, Borexino~\cite{Borexino:2019wln}, SK-I/II/III~\cite{Super-Kamiokande:2011lwo}, SK-IV~\cite{Super-Kamiokande:2021jaq},
    and KamLAND~\cite{KamLAND:2021gvi} for the monochromatic PBH mass function. The red solid curve is the combined constraint using all these experiments.}   \label{global}
\end{figure}

Fig.~\ref{global} shows the upper limits on the DM fraction of PBHs with a monochromatic mass function obtained using the upper limit on $\bar{\nu}_{e}$ fluxes data from current experiments as discussed in the previous section. 

It shows that in Fig.~\ref{global} (and can be expected from Fig.~\ref{fig:data}) that SNO has the weakest upper limits on $\bar{\nu}_{e}$ fluxes, resulting in a weak constraint on the fraction of PBH. For other detectors, Borexino can set the tightest upper limits for $\bar{\nu}_{e}$ fluxes at lower energies, even below $10\,{\rm MeV}$, due to its high energy resolution, low intrinsic backgrounds, and the low reactor $\bar{\nu}_{e}$ flux at the Gran Sasso site. In contrast, SK and KamLAND impose the strongest constraints above approximately $10 \,{\rm MeV}$, making them the most effective in constraining PBHs as 100\% of the DM candidate.
Compared to SK, KamLAND provides a better constraint at high mass because the neutron tagging efficiency of KamLAND provides an advantage over SK when searching in the lower neutrino energy region~\cite{KamLAND:2021gvi}. However, KamLAND's small fiducial volume limits its sensitivity compared to water-based Cherenkov detectors.

Given that each of these experiments operates independently and provides uncertainties at the same confidence level for each energy bin, it is possible to incorporate all the upper limits on $\bar{\nu}_{e}$ fluxes to establish a combined constraint. 
To achieve this, we modify the approach presented in Eq.~(\ref{eq:flux}) and consider the total $\chi^{2}_{tot} = \sum_{det} \chi^{2}_{det}$, where $\chi^{2}_{det}$ is the individual $\chi^{2}$ of the detectors. 
This enables us to combine the sensitivities of multiple experiments and obtain a more comprehensive assessment.

The combined limit is shown in Fig.~\ref{global} as the red line. It shows that PBHs alone cannot account for the entirety of DM up to masses of approximately $6.4 \times 10^{15}\,{\rm g}$. This signifies an enhancement of roughly 20\% compared to the
previous upper limits in the heaviest PBHs mass at $f_{\mathrm{PBH}}= 1$ obtained solely from data from SK~\cite{Dasgupta:2019cae}, which were around $5.2 \times 10^{15}\,{\rm g}$.

When the data from different experiments are combined statistically, it is important to recognize that these experiments may possess varying degrees of systematic uncertainties. A more careful analysis would need to cross calibrate the systematic differences of these detectors.

\subsection{Comparisons with existing results and future experimental prospects with neutrinos}

Fig.~\ref{fig:background} shows the results from our combined constraint and the existing results from SK~\cite{Dasgupta:2019cae}. Our results from the SK experiment deviate slightly from that reported in Ref.~\cite{Dasgupta:2019cae}. This discrepancy arises because the previous study utilized a $\bar{\nu}_{e}$ flux upper limit of 2.9 $ {\rm cm^{-2} s^{-1} }$ in the energy range of 17.3 to 19.3 MeV~\cite{Super-Kamiokande:2011lwo}, while our analysis, depicted by the green region in Fig.~\ref{global}, utilizes upper limits for different energy bins~\cite{Super-Kamiokande:2021jaq}. Additionally, our analysis takes into account the contribution of secondary neutrino emission, whereas the previous study only performed an analysis of primary emissions. While a more recent study~\cite{Bernal:2022swt} also included all spectral information of signal and background events shown as the yellow line, they considered not only the main detection channel of inverse beta decay but also the subdominant contributions from $\bar{\nu}_{e}$ and ${\nu}_{e}$ charged-current interactions off oxygen nuclei and $\bar{\nu}_{\mu}$ and ${\nu}_{\mu}$ charged-current interactions producing muons below the Cherenkov threshold. This effect is especially relevant for small PBH masses, which is responsible for their improved constraints over our results (and Ref.~\cite{Dasgupta:2019cae}) at small PBH masses.

For future prospects of searching for PBH with neutrinos, we consider the expected upper limits on $\bar{\nu}_{e}$ fluxes from JUNO~\cite{JUNO:2022lpc} and use the same procedure as discussed in Sec.~\ref{sec:chisq} to obtain
the expected constraint on the PBH fraction $f_{\mathrm{PBH}}$.  This is shown as the cyan dashed line in Fig.~\ref{fig:background}. Our results deviate slightly from those reported in Ref.~\cite{Wang:2020uvi} shown as the brown dashed line because they used a broader energy range, which produces stronger constraints at smaller PBH masses. 
The results from Ref.~\cite{Wang:2020uvi} are slightly stronger than those in Ref.~\cite{Bernal:2022swt}, which are shown by the blue dashed line. This may due to the use of integrated number of events and the higher neutrino fluxes~\cite{Bernal:2022swt}. 
In addition, the purple dashed line represents the expected limits from HK~\cite{Bernal:2022swt}, which are projected to be stronger than those from JUNO. We also present the expected limits from THEIA for two potential configurations: 25 and 100 kton, where the primary difference between the two THEIA configurations is approximately a factor of 2, attributed to the variance in statistics~\cite{DeRomeri:2021xgy}. Both of these future experimental prospects are expected to exclude PBHs as the sole component of DM up to masses of approximately $8 \times 10^{15}\,{\rm g}$.


\section{The Sensitivity Floor for Primordial Black Hole Neutrinos} \label{sec:floors}
\subsection{The DSNB as irreducible background}

Supernova neutrinos are produced during the gravitational collapse of the stellar core~\cite{Mirizzi:2015eza}. The emission of neutrinos from SN 1987A was first detected by the Kamiokande-II~\cite{Kamiokande-II:1987idp,Hirata:1988ad}, Irvine-Michigan-Brookhaven ~\cite{Bionta:1987qt, IMB:1988suc} and Baksan detectors~\cite{Alekseev:1988gp}. 
The DSNB is the collective flux of neutrinos emitted by all core-collapse supernovae that have occurred throughout the history of the Universe~\cite{Beacom:2010kk,Lunardini:2010ab,Suliga:2022ica}. The DSNB contributes to an isotropic flux of MeV neutrinos and encapsulates the combined supernova neutrino emission and their evolution with redshift.  
As neutrino experiments become more sensitive~\cite{Moller:2018kpn, JUNO:2015zny}, the DSNB could eventually become a background for PBHs neutrino searches (or any DM and exotic origin of neutrinos).

This is similar to the case in DM direct detection experiments~\cite{PandaX:2022aac,Schumann:2019eaa}, where coherent neutrino-nucleus scattering~\cite{Freedman:1973yd} would produce a recoil signal similar to that of DM-nucleon scattering. In this case, the theoretical lower limit of the DM cross section is sometimes referred as the ``neutrino floor" or ``neutrino fog"~\cite{Billard:2013qya,OHare:2021utq}. Below that, the search for DM parameters is background limited and becomes challenging. 

In our case, the DSNB would thus set a similar ``neutrino floor'' in the PBH search. While reactor neutrinos~\cite{Wurm:2007cy} and atmospheric neutrinos~\cite{Battistoni:2005yu} also contribute to the backgrounds, both fluxes could be measured accurately, and thus could be taken into account for the analysis. Although the DSNB has different expected angular and energy distribution compared to that of PBH neutrinos, the theoretical uncertainties in the DNSB flux means that once some ``signals'' due to the DSNB are detected, it will be difficult to uncover any subdominant PBH or DM components.  For concreteness, we consider HK as an example~\cite{Moller:2018kpn} for the PBH ``neutrino floor''.
 
\begin{figure}[t!]
    \centering
    \includegraphics[width=3.5in,height=3.0in]{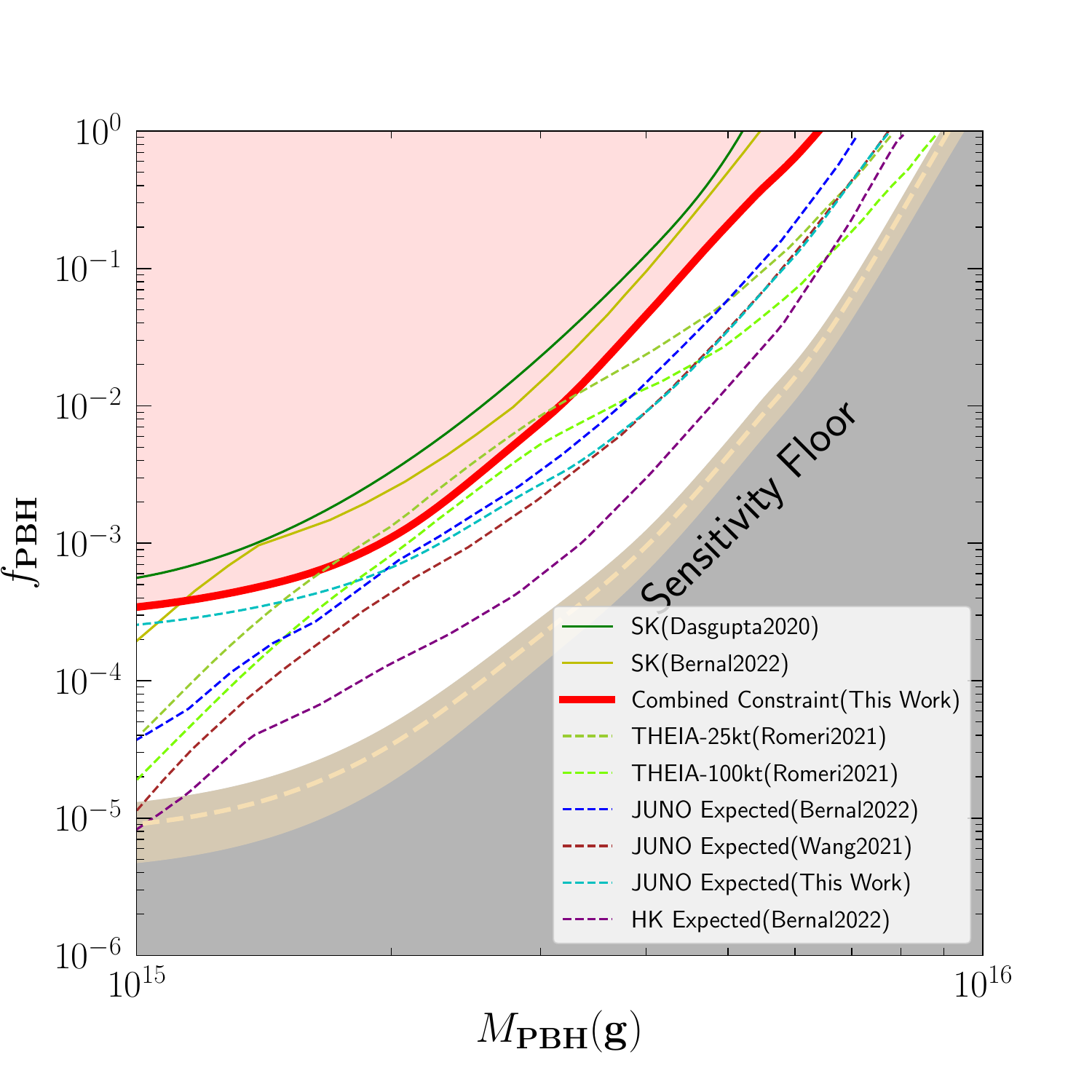}
    \caption{Current and future upper bounds on PBHs and the sensitivity floor for PBH neutrinos. The green solid line and the yellow solid line are the existing limits from SK~\cite{Dasgupta:2019cae,Bernal:2022swt}. Our combined constraints are shown in the red solid line.  We also show the expected constraints from JUNO~\cite{Bernal:2022swt,Wang:2020uvi}, THEIA~(25 and 100 kton)~\cite{DeRomeri:2021xgy},  and HK~\cite{Bernal:2022swt}.  The gray shaded region represents the sensitivity floor due to the DSNB background, while the yellow band accounts for the DSNB model uncertainty, and the dashed yellow line corresponds to the fiducial DSNB model~\cite{Moller:2018kpn}.}
    \label{fig:background}
\end{figure}

\subsection{Sensitivity floor for PBH neutrinos in HK}
To determine the sensitivity floor, we use the estimated DSNB event rate within the energy range of $E_{\nu} = 10\,{\rm MeV}$ to $30\,{\rm MeV}$ in next-generation neutrino detectors HK~\cite{Moller:2018kpn}.

The probability of observing $k_i$ counts in the energy bin $i$ can be described by the Poisson probability distribution function
\begin{equation}
P_i(k_{i} \mid \lambda_{i}) =  \frac{\lambda_{i}^{k_{i}} e^{-\lambda_{i}}}{k_{i} !}\, ,
\end{equation}
where $\lambda _i$ is the expected number of events.
Note that the model prediction, $\lambda _i$, depends on the model parameters $f_\mathrm{PBH}$ and $M_{\mathrm{PBH}}$. It is given by the sum of the PBH and the DSNB contributions, $\lambda_{i}=N^{\mathrm{PBH}}_i+N^{\mathrm{DSNB}}_i$.  The PBH contribution is 
\begin{equation}
N^{\mathrm{PBH}}_i=\varepsilon N_t t \int \frac{d \Phi^{\mathrm{PBH}}\left(E_{\nu}\right)}{d E_{\nu}} \sigma_i(E) d E,
\end{equation}
where $\varepsilon = 67\%$ is the detector efficiency, $N_t =  2.5 \times 10^{34}$ is the number of targets, $ \sigma_i(E)$ corresponds to the IBD cross section~\cite{Strumia:2003zx,Ricciardi:2022pru,Ricciardi:2023tvg}, and $t=20~\mathrm{yrs}$ represents the total data taking time~\cite{Moller:2018kpn,Tabrizi:2020vmo}.

 Similarly, the number of expected events from the DSNB at the $i$th energy is given by
\begin{equation}
N^{\mathrm{DSNB}}_i=\varepsilon N_t t \int \frac{d \Phi^{\mathrm{DSNB}}\left(E_{\nu}\right)}{d E_{\nu}} \sigma_i(E) d E,
\end{equation}
 where $\frac{d \Phi^{\mathrm{DSNB}}\left(E_{\nu}\right)}{d E_{\nu}}$ is the predicted DSNB flux as a function of the neutrino energy using so-called “fiducial DSNB model”~\cite{Moller:2018kpn}.

The joint likelihood of observing $N$ energy bins is given by
\begin{equation}
{L} =\prod_{i=1}^{N} P_i(k_{i} \mid \lambda_{i}) = \prod_{i=1}^{N} \frac{\lambda_{i}^{k_{i}} e^{-\lambda_{i}}}{k_{i} !}.
\end{equation}
Here, we assume that the observed number of events $k_i$ is given by the background only, i.e., $k_{i} = N^{\mathrm{DSNB}}_{i}$.  We note that only $N^{\mathrm{PBH}}_i$ depends on $f_{\rm PBH}$ and that $\lambda_{i}(f_{\rm PBH} = 0) = k_{i}$.

To obtain the expected sensitivity, we use the log-likelihood ratio~\cite{Cowan:2010js}
\begin{equation}
TS =-2 \ln \frac{L\left(f_{\mathrm{PBH}}\right)}{L\left(f_{\mathrm{PBH}} = 0\right )}.
\end{equation}
The limit on the PBH fraction at a one-sided 95\% confidence level can be determined by $TS =2.71.$
 
Fig.~\ref{fig:background}, in the gray region, shows the sensitivity floor for PBHs with neutrinos. It is worth noting that PBHs with masses exceeding $9 \times 10^{15}\,{\rm g}$ would be challenging to be detected by neutrinos from Hawking radiation. To account for the theoretical uncertainties inherent in the DSNB model, we consider its range of variability~\cite{Moller:2018kpn}, represented by a yellow band in Fig.~\ref{fig:background}.

We note that the definition of the sensitivity floor is not unique~\cite{Billard:2013qya,Edsjo:2017kjk,Arguelles:2017eao}. Therefore, it is also possible to define a theoretical sensitivity floor such that the neutrino flux from PBHs equals the flux from the DSNB at some energies.
This approach would be detector independent. 
However, we find that the theoretical sensitivity floor defined in this manner tends to be higher than the detector dependent approach. 
This is because the events approach takes into the Poisson events probabilities, while the theoretical flux approach relies on the flux comparison at a single point.
Thus, the detector dependent definition of the sensitivity floor based on number of events is more practical and is presented in this work.

\section{Conclusions and discussion}
 
In this work, we investigate the constraints on the abundance of nonrotating PBHs with monochromatic mass distributions ranging from $5 \times 10^{14}\,{\rm g}$ to $10^{16}\,{\rm g}$ by utilizing the upper limits on $\bar{\nu}_{e}$ fluxes. 
To establish constraints on the abundance of PBHs, we consider data from several experiments such as SNO~\cite{SNO:2004eru}, Borexino~\cite{Borexino:2019wln}, SK-I/II/III~\cite{Super-Kamiokande:2011lwo,Super-Kamiokande:2013ufi}, SK-IV~\cite{Super-Kamiokande:2021jaq}, and KamLAND~\cite{KamLAND:2021gvi}.

By incorporating the data from all available experiments, we improve upon the constraints on the PBHs abundance, ruling out the possibility of them being the sole component of DM for masses up to approximately $6.4 \times 10^{15}\,{\rm g}$. This represents an improvement of approximately 20\% in comparison to the previous upper limits in the heaviest PBHs mass at $f_{\mathrm{PBH}}= 1$ obtained from SK's data~\cite{Dasgupta:2019cae}. Notably, the upcoming JUNO experiment and HK could potentially extend the constraints on PBHs up to masses of approximately $8 \times 10^{15}\,{\rm g}$. 

Finally, we evaluate the sensitivity floor for PBH neutrinos due to the DSNB using the expected data from HK and find that PBHs with masses higher than $9 \times 10^{15}\,{\rm g}$ would be difficult to detect due to the presence of the DSNB.  Thus, there is a limited parameter space (as shown in Fig.~\ref{fig:background}) that could be probed using neutrinos. To probe PBH as 100\% of the DM candidate beyond this mass range, other messengers, such as electrons/positrons and gamma rays are needed~\cite{Laha:2020ivk,Ballesteros:2019exr,Arbey:2019vqx,Coogan:2020tuf,Carr:2009jm,Ray:2021mxu,Dasgupta:2019cae,Laha:2019ssq,Chen:2021ngo,Tan:2022lbm,Malyshev:2022uju}. 
\bigskip
\section*{Acknowledgments}
We are grateful for the helpful discussions with Anupam Ray, Ranjan Laha and, Chingam Fong. The works of Q. L. and K. C. Y. N. are supported by Croucher foundation, RGC Grants No. 24302721, No. 14305822, and No. 14308023, and NSFC/GRC Grant No. N\_CUHK456/22.
\bibliography{bib.bib}
\end{document}